\pdfoutput=1

\documentclass[11pt,twoside,a4paper,cmspaper,final,collab]{cms-tdr}
\def\svnVersion{4b532cf}\def\svnDate{2020/11/25}\def\cmsCernNoTag{CERN-EP-2020-101}\def\cmsCernDate{\today}\def\cmsMessage{"Published in Physical Review Letters as \href{http://dx.doi.org/10.1103/PhysRevLett.125.222001}{\doi{10.1103/PhysRevLett.125.222001}.}"}

\begin{document}\cmsNoteHeader{HIN-19-001}

\hyphenation{had-ron-i-za-tion}
\hyphenation{cal-or-i-me-ter}
\hyphenation{de-vices}
\newcommand{\roots} {\ensuremath{\sqrt{\smash[b]{s}}}\xspace}
\newcommand{\PbPb}{\ensuremath{\mathrm{Pb}\mathrm{Pb}}\xspace}
\newcommand{\pPb}{\ensuremath{\Pp\mathrm{Pb}}\xspace}
\newcommand{\pp}{\ensuremath{\Pp\Pp}\xspace}
\newcommand{\pn}{\ensuremath{\Pp\Pn}\xspace}
\newcommand{\ppbar}{\ensuremath{\Pp\PAp}\xspace}
\newcommand{\jj}{\ensuremath{ \mathrm{j} \mathrm{j}^{\prime}}\xspace}
\newcommand{\mjj}{\ensuremath{m_{\jj}}\xspace}
\newcommand{\mllpmmp}{\ensuremath{m(\ell^\pm \ell^\mp)}\xspace}
\newcommand{\mllpm}{\ensuremath{m(\ell^+ \ell^-)}\xspace}
\newcommand{\mtop}{\ensuremath{_{\cPqt}}\xspace}
\newcommand{\stt}{\ensuremath{\sigma_{\ttbar}}\xspace}
\newcommand{\DeliLumi}{\ensuremath{ 1.8\nbinv}\xspace}
\newcommand{\RecoLumi}{\ensuremath{ \text{(}1.7\pm 0.1\text{)}\nbinv}}
\newcommand{\empm}{\ensuremath{\Pepm{}\Pgm^\mp}}
\newcommand{\mmpm}{\ensuremath{\PGmp{}\PGmm{}}}
\newcommand{\empe}{\ensuremath{\Pep{}\Pem{}}}
\newcommand{\eb}{\ensuremath{\varepsilon_{\PQb}}}
\newcommand{\methone}{dilepton-only}
\newcommand{\methtwo}{dilepton plus \PQb-tagged jets}
\newcommand{\methonecap}{Dilepton only}
\newcommand{\methtwocap}{Dilepton plus \PQb-tagged jets}
\newcommand{\mub}{\ensuremath{\,\mu\text{b}}\xspace}
\newcommand{\nb}{\ensuremath{N_{\operatorname{\PQb\text{-}tag}}}}
\newcommand{\Zgstar}{\ensuremath{\cPZ/\PGg^{*}}}

\newcommand{\zvalexptwol}{\ensuremath{ 4.8}}
\newcommand{\zvalobstwol}{\ensuremath{ 3.8}}
\newcommand{\zvalexptwoljets}{\ensuremath{ 5.8}}
\newcommand{\zvalobstwoljets}{\ensuremath{ 4.0}}
\newcommand{\muexptwol}{\ensuremath{ 1.00^{+0.25}_{-0.23}}}
\newcommand{\muobstwol}{\ensuremath{ 0.79^{+0.26}_{-0.23}}}
\newcommand{\muexptwoljets}{\ensuremath{ 1.00^{+0.23}_{-0.21}}}
\newcommand{\muobstwoljets}{\ensuremath{ 0.63^{+0.22}_{-0.20}}}
\newcommand{\stttwol}{\ensuremath{2.54^{+0.84}_{-0.74}}}
\newcommand{\stttwoljets}{\ensuremath{2.03^{+0.71}_{-0.64}}}

\ifthenelse{\boolean{cms@external}}{\providecommand{\cmsLeft}{upper\xspace}}{\providecommand{\cmsLeft}{left\xspace}}
\ifthenelse{\boolean{cms@external}}{\providecommand{\cmsRight}{lower\xspace}}{\providecommand{\cmsRight}{right\xspace}}
\ifthenelse{\boolean{cms@external}}{\providecommand{\suppMaterial}{the supplemental material~\cite{supp}}}{\providecommand{\suppMaterial}{Appendix~\ref{app:suppMat}}}

\cmsNoteHeader{HIN-19-001}
\title{Evidence for top quark production in nucleus-nucleus collisions}

\date{\today}

\abstract{
Ultrarelativistic heavy ion collisions  recreate in the laboratory the thermodynamical conditions prevailing in the early universe up to $10^{-6}$ seconds, thereby allowing the study of the quark-gluon plasma (QGP), a state of quantum chromodynamics (QCD) matter with deconfined partons. The top quark, the heaviest elementary particle known, is accessible in nucleus-nucleus collisions at the CERN LHC, and constitutes a novel probe of the QGP. Here, we report the first evidence for the production of top quarks in nucleus-nucleus collisions, using lead-lead collision data at a nucleon-nucleon center-of-mass energy of 5.02\TeV recorded by the CMS experiment. Two methods are used to measure the cross section for top quark pair production (\stt) via the selection of charged leptons (electrons or muons) and bottom quarks. One method relies on the leptonic information alone, and the second one exploits, in addition, the presence of bottom quarks. The measured cross sections, $\stt = \stttwol$ and $\stttwoljets$\mub, respectively, are compatible with expectations from scaled proton-proton data and QCD predictions.
}
\hypersetup{%
pdfauthor={CMS Collaboration},%
pdftitle={Evidence for top quark production in nucleus-nucleus collisions},%
pdfsubject={CMS},%
pdfkeywords={CMS, physics, heavy ions, top quark}}

\maketitle

Since the top quark discovery at the Fermilab Tevatron more than twenty years ago~\cite{Abe:1995hr,D0:1995jca},
top quarks have been measured at the LHC in proton-proton
(\pp)~\cite{Aad:2019hzw,Aaij:2018imy,Sirunyan:2018goh,Sirunyan:2017ule,Khachatryan:2015fwh}
as well as  proton-nucleus~\cite{Sirunyan:2017xku} collisions, but so far have not been observed
in nucleus-nucleus collisions because of insufficient nucleon-nucleon ($\mathrm{NN}$) center-of-mass energies (\sqrtsNN) or integrated
luminosities. The multi-\TeV energies available at the CERN LHC have opened up the possibility
to measure, for the first time, the top quark in lead-lead (\PbPb)
collisions~\cite{dEnterria:2015mgr}. More specifically, the top quark constitutes a novel and theoretically precise probe
of the nuclear parton distribution functions (nuclear PDFs, or nPDFs), in the poorly explored region where partons have a large fraction of the nucleon momentum, as well as of the
properties of the produced quark-gluon plasma (QGP)~\cite{dEnterria:2015mgr,Apolinario:2017sob}.
First, precise knowledge of nPDFs is a key prerequisite to extract 
detailed information on the QGP properties from the experimental data. Second, top quarks, on average, decay on a timescale similar to the formation of the QGP, hence offering a unique opportunity to study its time evolution~\cite{Apolinario:2017sob}.
The study presented here shows evidence for the production of the top quark in \PbPb\ collisions at $\sqrtsNN = 5.02\TeV$
with an integrated luminosity of \RecoLumi~\cite{CMS-PAS-LUM-17-002} as recorded by the CMS detector~\cite{Chatrchyan:2008zzk}.

The top quark---the heaviest elementary particle known---is produced at hadron colliders predominantly in pairs (\ttbar) through quantum chromodynamics (QCD)
processes, mostly gluon-gluon fusion at the LHC, and is thereby a sensitive probe of the gluon PDF 
of the incoming nucleons~\cite{Czakon:2013tha}. Once produced, it decays with almost 100\% probability
into a \PW boson and a bottom (\PQb) quark. Top quark pair production is thereby characterized by final states comprising the decay products of the two \PW{} bosons, and two \PQb{} jets, resulting from the hadronization products of \PQb{} quarks. The dilepton final states, in which both \PW{} bosons decay into
electrons (\Pe) or muons (\Pgm) and the corresponding neutrinos ($\nu$),
are the cleanest final states for the \ttbar signal measurement, despite their relatively small branching fraction
$\mathcal{B}(\ttbar \to \ell^+\ell^-\,\nu_\ell\PAGn_\ell\,\PQb\PAQb)=5.25\%$~\cite{Tanabashi:2018oca}, with $\ell^\pm = \Pe^\pm, \Pgm^\pm$.
See \suppMaterial{} for a candidate \ttbar\ event.

In this letter, the measurement of the \ttbar\ cross section (\stt) in three dilepton final states, \ie, \empe, \mmpm, and \empm, is performed by (i) making use of the final-state
dilepton kinematic properties alone, and (ii) imposing extra requirements on the number of jets
``tagged'' as originating from \PQb quarks (referred to as ``\PQb-tagged jets'').
The first method is motivated by the fact that leptons propagate unscathed 
through the QGP, thereby providing favorable conditions for the detection of \ttbar production.
The second method, which enhances the signal over background in standard \pp analyses, is applied
with realistic estimates of the impact of \PQb quark energy loss, also known as ``jet quenching'', in the QGP~\cite{Chatrchyan:2013exa}.

The main feature of the CMS detector is a superconducting solenoid, providing a magnetic field of 3.8\unit{T}.
Within the solenoid volume is a silicon pixel and strip tracker, which is used to detect charged particles,
a lead tungstate crystal electromagnetic calorimeter, and a brass and scintillator hadron calorimeter.
Hadron forward calorimeters extend the pseudorapidity coverage up to $\abs{\eta}=5.2$.
Muons are detected in gas-ionization chambers embedded in the steel flux-return yoke outside the solenoid.
A more detailed description of the CMS detector, together with a definition of the coordinate system, the relevant kinematic variables (\eg, the transverse momentum $\pt$),
and the physics-object reconstruction, can be found in Ref.~\cite{Chatrchyan:2008zzk}.
The data sample is collected with a two-level trigger system~\cite{Khachatryan:2016bia}:
at level-1 events are selected by custom hardware processors while the high-level trigger uses fast versions of the offline software.
All particle candidates are reconstructed with the particle-flow algorithm~\cite{Sirunyan:2017ulk} using an optimized combination of information from the various elements of the CMS detector.

The data sample is filtered to favor events with two opposite-sign (OS) high-\pt leptons, with $\pt > 25$ (20)\GeV and $\abs{\eta} < 2.1$ ($2.4$)
for the electron~\cite{Khachatryan:2015hwa} (muon~\cite{Sirunyan:2018fpa}) candidates, that do not belong to jets reconstructed using a distance parameter of 0.4~\cite{Cacciari:2011ma,Cacciari:2008gp}
and are thus isolated from nearby hadronic activity. The correction for the underlying event (UE) formed from soft processes is done on a particle-by-particle basis using an estimation of the median of the energy density $\rho$~\cite{Cacciari:2008gp,Cacciari:2007fd,Cacciari:2011ma} in the event for the lepton isolation, and the ``constituent subtraction'' method~\cite{Berta:2014eza,Sirunyan:2017bsd} for the jet constituents. The characteristic additional presence of two \PQb-tagged jets in the $\ttbar$ decay chain is then used, in our second method only, to enhance the sensitivity 
to the top quark signal. Jets must have $\pt> 30\GeV$ and $\abs{\eta}< 2.0$, and are considered as \PQb tagged if an optimized ``combined secondary vertex'' (CSV) discriminator~\cite{Sirunyan:2017ezt} produces a value above a certain threshold for the probability of the jet to stem from the hadronization of the \PQb quark.
The \PQb tagging efficiency depends on the geometric overlap of the two colliding nuclei as given by the centrality percentile~\cite{Chatrchyan:2012nia}. This percentile is defined from percentages corresponding to fractions of the total inelastic hadronic cross section, \eg, with 0\% denoting the full overlap of the two colliding nuclei. 
After the selection criteria, the \PQb tagging efficiency in a $\mathrm{NN} \to \ttbar$ Monte Carlo (MC) simulation sample, generated at next-to-leading order (NLO) in QCD using the \MGvATNLO(v2.4.2)~~\cite{Alwall:2014hca} program and interfaced to the ``tuned''~\cite{Sirunyan:2019dfx} \PYTHIA{}8~(v2.3.0)~\cite{Sjostrand:2014zea} MC event generator, is approximately 60\,(70)\%,
with a misidentification rate of 5\,(2)\%, in the 0--30  (30--100)\% centrality interval.
The two jets with the highest CSV discriminator values are used to count the \PQb-tagged jet multiplicity, \nb, and classify the selected events into the ``0\PQb{}'' ($\nb=0$), ``1\PQb{}'' 
($\nb=1$), and ``2\PQb{}'' ($\nb=2$) jet categories.

The main background is Drell--Yan quark-antiquark annihilation into lepton-antilepton pairs through \cPZ bosons or virtual photons
(a process referred to as ``\Zgstar''). It contaminates all final states with either offshell (in \empe and \mmpm) or $\Zgstar \to \Pgt^+\Pgt^- \to \Pe^\pm\Pgm^\mp+X$ (in \empm) decays, where ``$X$'' represents other particles. The \Zgstar\ process is modeled at NLO using the \MGvATNLO simulation with corrections obtained from data, as detailed below.
In the \empm\ final state, in particular, there are  additional contaminations from \PW\ boson production 
in association with jets (``\PW{}+jets''), \Zgstar\ with one unreconstructed lepton, 
and QCD multijet events. In these cases, the produced jets are mainly from heavy quarks eventually
decaying into high-\pt leptons that are erroneously identified as being isolated. These latter processes, referred to in what follows as
``nonprompt'' background, are directly derived from control regions in the data, as explained next.
Smaller background contributions from single top quark and \PW{} boson (``$\PQt\PW$''), and \PW{}\PW, \PW{}\cPZ, and \cPZ{}\cPZ\ 
(collectively referred to as ``$\PV\PV$'') production, are directly estimated from MC simulation with \POWHEG~\cite{Re:2010bp,Nason:2013ydw}.
In all simulated samples, the EPPS16 NLO nPDF~\cite{Eskola:2016oht}, with  CT14 NLO free-nucleon PDF~\cite{Dulat:2015mca},
the strong coupling constant at the \cPZ boson mass $\alpS(m_{\cPZ})=0.118 \pm 0.001$~\cite{Tanabashi:2018oca},
and the top quark mass $m_{\cPqt}=172.5\GeV$~\cite{Sirunyan:2018mlv} are used as input. At the step of detector digitization, each hard scattering event is placed at the same primary-vertex~\cite{CMSCollaboration:2015zni} location as a heavy ion background event generated with \HYDJET~(v1.9)~\cite{Lokhtin:2005px}, to mimic the effects of
the UE without any QGP-induced modifications of the final-state particles from the top quark decay.
Finally, all simulated samples include an emulation of the full CMS detector response, based on \GEANTfour~\cite{Agostinelli:2002hh}, and a realistic description of the luminous region produced by the collisions.

The \Zgstar\ simulation provides a good modeling of the dilepton kinematic properties, except for the low-\pt region where multiple soft-gluon emission dominates and the agreement is slightly worse.
We thus apply correction (``scale'') factors to the MC simulation using events in data enriched with $\Zgstar \to \ell^+\ell^-$ candidates.
The scale factors are measured as a function of centrality, but no particular centrality dependence is seen.
The difference between the corrected and uncorrected MC distributions is considered as the \Zgstar\ \pt modeling uncertainty.
Events in the \empe\ and \mmpm\ final states with dilepton invariant mass \mllpm in the proximity of the \cPZ boson mass $m_{\cPZ}$~\cite{Tanabashi:2018oca} ($76<\mllpm <106 \GeV$) are rejected, and their number is used to control the normalization of the corrected MC distributions outside the $m_{\cPZ}$ region.
The overall normalization of the nonprompt background is estimated by forming a ``same-sign'' (SS) control region, \ie, applying the same criteria as to the signal selection, but requiring SS lepton
pairs. The SS dilepton events predominantly contain at least one misidentified lepton.
The scaling from the SS control to the signal regions is performed assuming the ratio of the number of OS to SS events containing misidentified leptons
to be unity. To estimate the distribution of the nonprompt background, an event mixing technique is developed. The mixing is performed for each lepton in a pool of 100 different events sharing the same features (\ie, lepton charge and flavor, and whether originating from onshell or offshell \cPZ bosons). Each lepton is randomly substituted, and the kinematic variables are recomputed with this new dilepton hypothesis. A multidimensional distance is calculated with respect to the original event using a nearest-neighbor algorithm~\cite{Pedregosa:2012toh}. The variables entering the algorithm are  the centrality, $\rho$, lepton \pt and isolation, and the magnitude of the \pt of the dilepton system (``dilepton \pt''). The highest ranked mixed events, corresponding to the smallest multidimensional distance, are chosen as the nominal distribution. Differences with respect to the distributions obtained using events further apart in this multidimensional distance, \ie, lower ranked hypotheses, are considered as a source of systematic uncertainty.

For both the \methone\ and \methtwo\ methods, a boosted decision tree (BDT~\cite{Hocker:2007ht}) classifier is trained on the simulated \ttbar signal versus the overall
$\Zgstar \to \empe$,$\mmpm$ background. This classifier is based exclusively on leptonic quantities to minimize effects from the imprecise knowledge of the jet
properties in the heavy ion environment. The BDT exploits the properties of the leading- and subleading-\pt leptons, denoted by  ``$\ell_1$'' and ``$\ell_2$'', respectively, and their correlations. As input to the BDT classifier, the following variables are used in descending order of importance:
(i)~the \pt of the leading lepton, $\pt (\ell_1)$,
(ii)~the normalized momentum imbalance between $\ell_1$ and $\ell_2$, $A_{\pt}=\left(\pt (\ell_1) - \pt (\ell_2)\right)/\left(\pt (\ell_1) + \pt (\ell_2)\right)$,
(iii)~the dilepton \pt,
(iv)~the absolute pseudorapidity of the dilepton system,
(v)~the absolute azimuthal separation between $\ell_1$ and $\ell_2$, and
(vi)~the sum of the absolute $\eta$ of $\ell_1$ and $\ell_2$.

\begin{figure}[!htbp]
    \centering
    \includegraphics[width=0.49\textwidth]{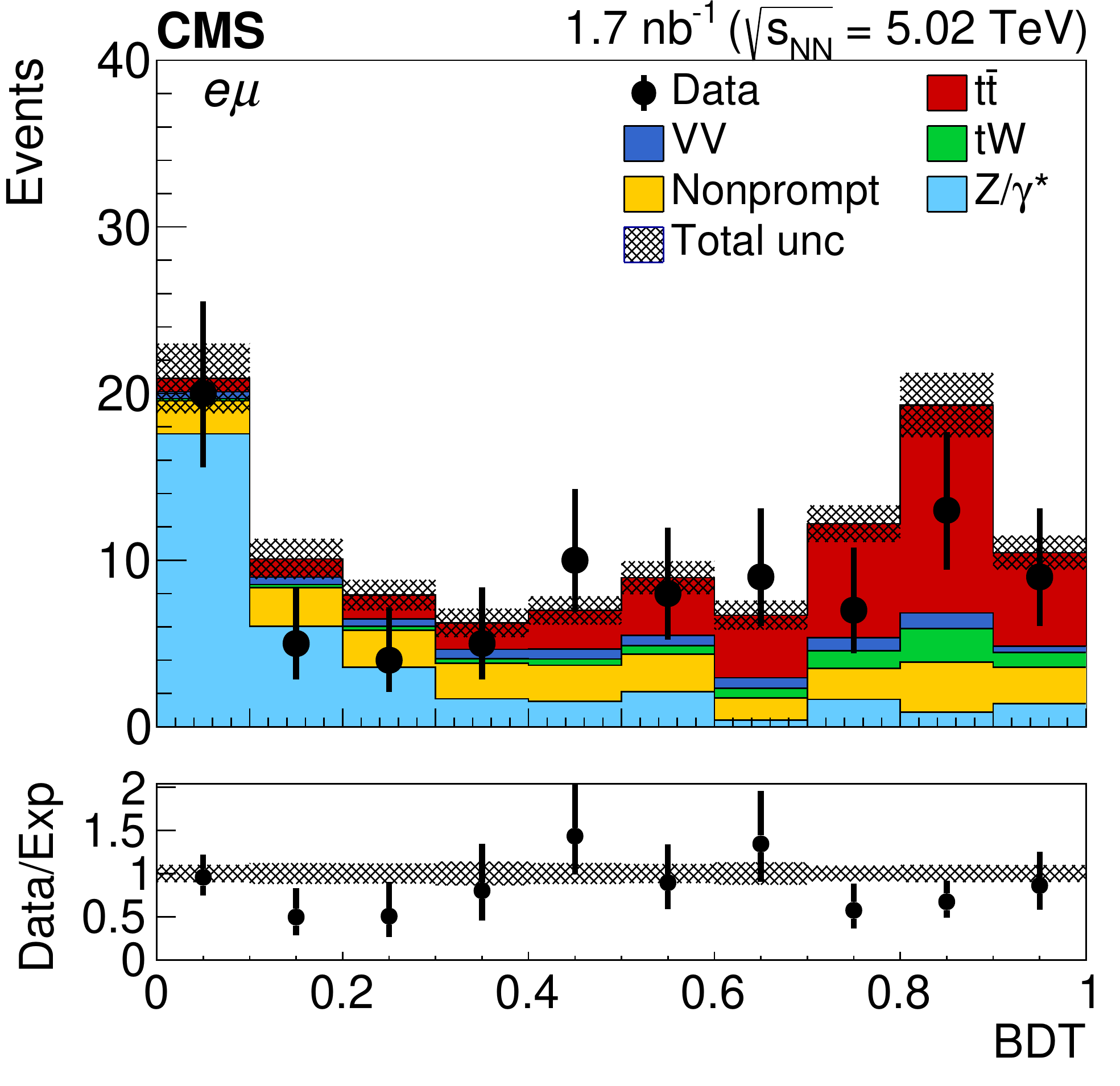}
    \includegraphics[width=0.49\textwidth]{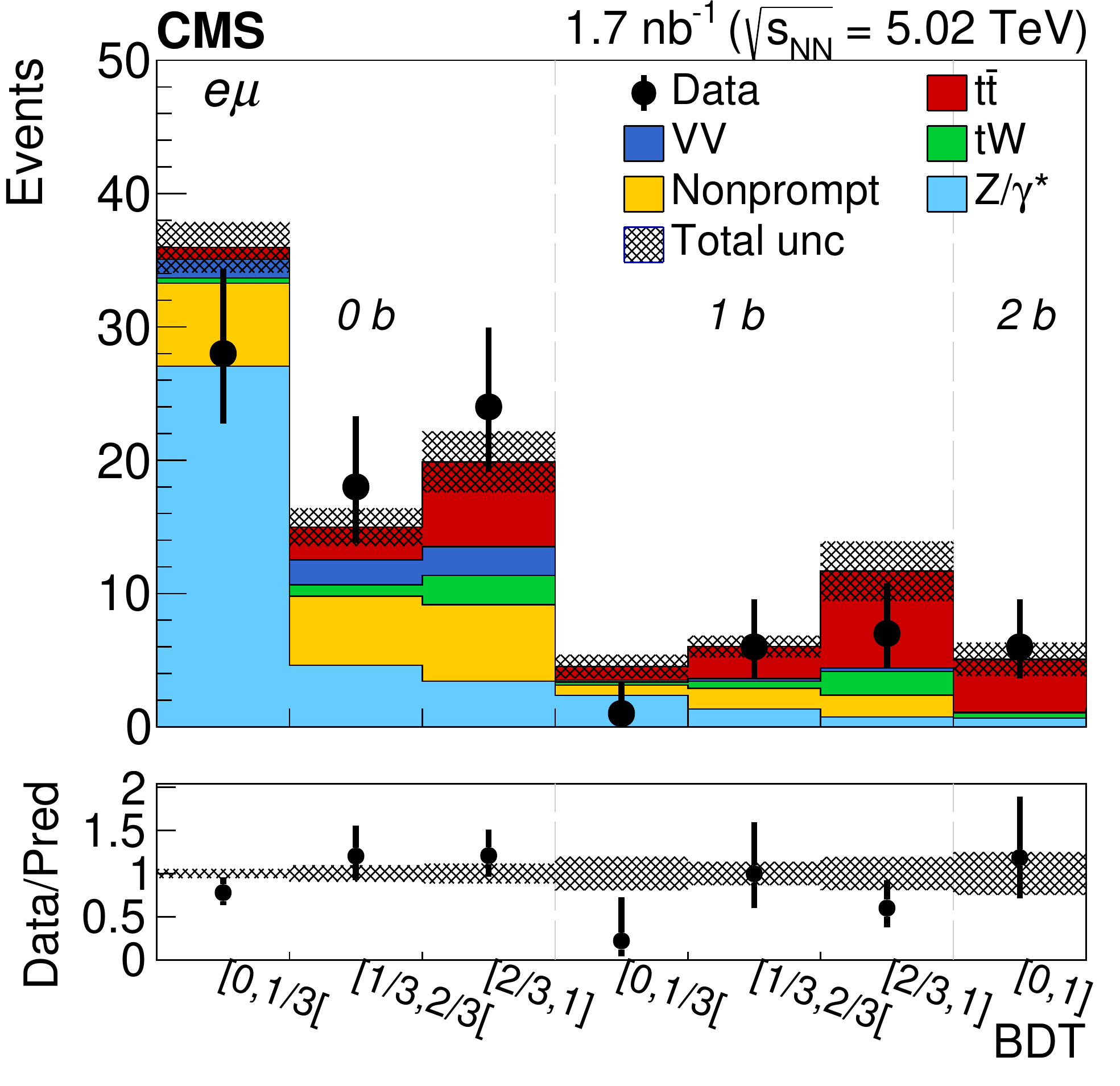}
    \caption{Observed and prefit expected (left) or postfit predicted (right) BDT discriminator distributions in the \empm\ final state
    either inclusively (\cmsLeft) or separately in the 0\PQb-, 1\PQb-, and 2\PQb-tagged jet multiplicity categories (\cmsRight).
    The data are shown with markers, and the signal and background processes with filled histograms.
    The vertical bars on the markers represent the statistical uncertainties in data. The hatched regions show the uncertainties in the sum of \ttbar\ signal and backgrounds.
    The lower panels display the ratio of the data to predictions, including the \ttbar\ signal,  with bands representing the uncertainties in the predictions.
      }
    \label{fig:resultsBDT}
\end{figure}

Figure~\ref{fig:resultsBDT} (\cmsLeft) shows the observed BDT discriminator distribution for the \methone\ method in the higher sensitivity \empm\ final state
(see \suppMaterial{} for the \empe and \mmpm\ final states). The \ttbar\ signal and various sources of background are also shown, indicated as ``prefit expected'' as they are not adjusted according to any statistical treatment (``fit''). The  \Zgstar background is normalized to the next-to-next-to-leading order (NNLO) cross section from the \FEWZ (v3.1.rc) program~\cite{Li:2012wna},
and the $\PV\PV$ and $\PQt\PW$ contributions are normalized to the NLO and the approximate NNLO cross sections calculated with \MCFM~(v8.0)~\cite{Campbell:2011bn} and from Ref.~\cite{Kidonakis:2013zqa}, respectively. The classifier separates well the \ttbar signal from the \Zgstar\ background in all final states.
The \ttbar signal (red histogram) populates the high-BDT discriminator values.
The uncertainties in the data are statistical only, while the uncertainties in the backgrounds include a prefit expectation of the systematic
uncertainty.

Profile likelihood fits~\cite{Verkerke:2005,Moneta:2010pm} to binned BDT discriminator distributions are performed
separately for the \methone\ and \methtwo\ methods. The best fit values of ``signal strength'' ($\mu$), their uncertainty $\Delta\mu$
(corresponding to a 68\% confidence level), and the significance (in units of standard deviations) of the \ttbar process against the background-only hypothesis,
are obtained following the procedure described in Section 3.2 of Ref.~\cite{Khachatryan:2014jba} and the frequentist paradigm using the profile likelihood ratio as a test statistic~\cite{Cowan:2010js}, accordingly. The value of $\mu$ is defined as the ratio of the observed \stt\ to the expectation from theory, \ie, $\mu=\stt/\stt^{\text{th}}$.
The theoretical cross section  $\stt^{\text{th}} = \sigma_{\PbPb \to \ttbar+X}^{\mathrm{NNLO+NNLL}} = 3.22\, ^{+0.38}_{-0.35}\,(\text{nPDF}\oplus\text{PDF})\,^{+0.09}_{-0.10}\,\text{(scale)}$\mub,
calculated with the \textsc{Top++}~(v2.0) program~\cite{Czakon:2011xx, Czakon:2013goa} at NNLO in QCD, including soft-gluon resummation at next-to-next-to-leading logarithmic (NNLL) accuracy,
with the nuclear EPPS16~\cite{Eskola:2016oht} and free-nucleon CT14~\cite{Dulat:2015mca} PDFs. The same calculation but with the free-nucleon CT14 and NNPDF30~\cite{Ball:2014} NNLO PDFs (scaled by the square of the number of nucleons in the $\mathrm{Pb}$ nucleus, $A^2 = 208^2$) yields 
$\sigma_{\PbPb \to \ttbar+X}^{\mathrm{NNLO+NNLL}} = 3.04\,^{+0.18}_{-0.14}\,(\text{PDF})\,^{+0.08}_{-0.10}\,\text{(scale)}$
and $2.98 \pm 0.14\,(\text{PDF}\oplus\alpS(m_{\cPZ}))\,^{+0.08}_{-0.10}\,\text{(scale)}$\mub, respectively. 
The small difference between the cross sections obtained with nuclear and free-nucleon PDFs arises from the nPDF ``antishadowing'' effect, which is only mildly dependent on \sqrtsNN~\cite{dEnterria:2015mgr}.
 
In the \methone\ method, the extracted values of $\mu = \muobstwol$ (where contributions to $\Delta\mu$ are statistical and systematic in nature)
and significance of \zvalobstwol\ standard deviations constitute the first evidence of \ttbar production in nucleus-nucleus collisions.
As can be seen in the prefit distribution of Fig.~\ref{fig:resultsBDT} (\cmsLeft), the data are somewhat below the expectation at the high-BDT discriminator values.
This is also reflected in differences in $\mu$ and significance, where the expected values are $\mu = \muexptwol$ and \zvalexptwol\ standard deviations, respectively.

Events in which $\nb\geq 1$ are expected to be very pure in the \ttbar\ signal process.
Since \PQb quarks are affected by final-state energy loss in the QGP, 
we take into account the centrality-dependent impact from jet quenching on \nb.
We make use of a jet quenching model~\cite{Arleo:2002kh, Arleo:2017ntr} that is consistent with the CMS \PQb jet data~\cite{Chatrchyan:2013exa}, estimating the expected migration of \ttbar signal events among the 0\PQb{}-, 1\PQb{}-, or 2\PQb-tagged jet categories.
A combined profile likelihood fit, introducing a parameter \eb\ that correlates the number of \ttbar signal events in the three \PQb-tagged jet categories
based on multinomial probabilities~\cite{Sirunyan:2018goh},
 is thus expected to control better the background contamination.
We include in the likelihood  the effects on \eb\ from jet quenching (comparing the maximum with no \PQb quark energy loss scenarios), and the intrinsic uncertainties in the \PQb tagging efficiency and misidentification rate.
The values of the observed (expected) signal strength and significance are $\mu = \muobstwoljets$
($\muexptwoljets$) and  \zvalobstwoljets\ (\zvalexptwoljets) standard deviations, respectively.
Figure~\ref{fig:resultsBDT} (\cmsRight) compares the data to the \ttbar\ signal and various sources of background adjusted according to the fit procedure  (``postfit predicted'') for the \methtwo\ method in the \empm\ final state (see \suppMaterial{} for the \empe and \mmpm\ final states).
The BDT distribution for the \Zgstar\ background is taken from the MC simulation, after scaling the event yield in each \nb\
bin to the corresponding \nb\ distribution observed in data within the $m_{\cPZ}$ region.

Sources of experimental uncertainties, incorporated into the likelihoods via ``nuisance parameters'', include the lepton selection efficiency found using a ``tag-and-probe'' method~\cite{Chatrchyan:2013iaa} (6\%), integrated luminosity~\cite{CMS-PAS-LUM-17-002} (5\%), and the normalization of the background based on control samples in data (12\%). The statistical uncertainties  in  the \ttbar\
signal and background distributions (7\%) are estimated separately.
The \methtwo\ method is, in addition, affected by the uncertainty in \eb\ (6\%), and the jet energy scale and resolution (2\%) estimated following the methodology of Ref.~\cite{Khachatryan:2016kdb}.
Sources of theoretical uncertainty included in the likelihoods are
the nPDF parametrization (derived from the $54{+}40$ eigenvalues of the $\text{EPPS16}{+}\text{CT14}$ sets), the choice of renormalization and factorization scales
(within a factor of two from their default values of $\mu_\text{R}=\mu_\text{F}=m_{\cPqt}$), and $\alpS(m_{\PZ})$ are included ($<$1\%).
We also take into account the uncertainties in the \pt modeling of the \ttbar\ signal and \Zgstar\ background distributions (5\%)
as well as in the top quark mass ($<$1\%).
The precision of the two methods is dominated by the statistical uncertainty (${\approx}28\%$).

\begin{figure}[htbp]
\centering
\includegraphics[width=0.49\textwidth]{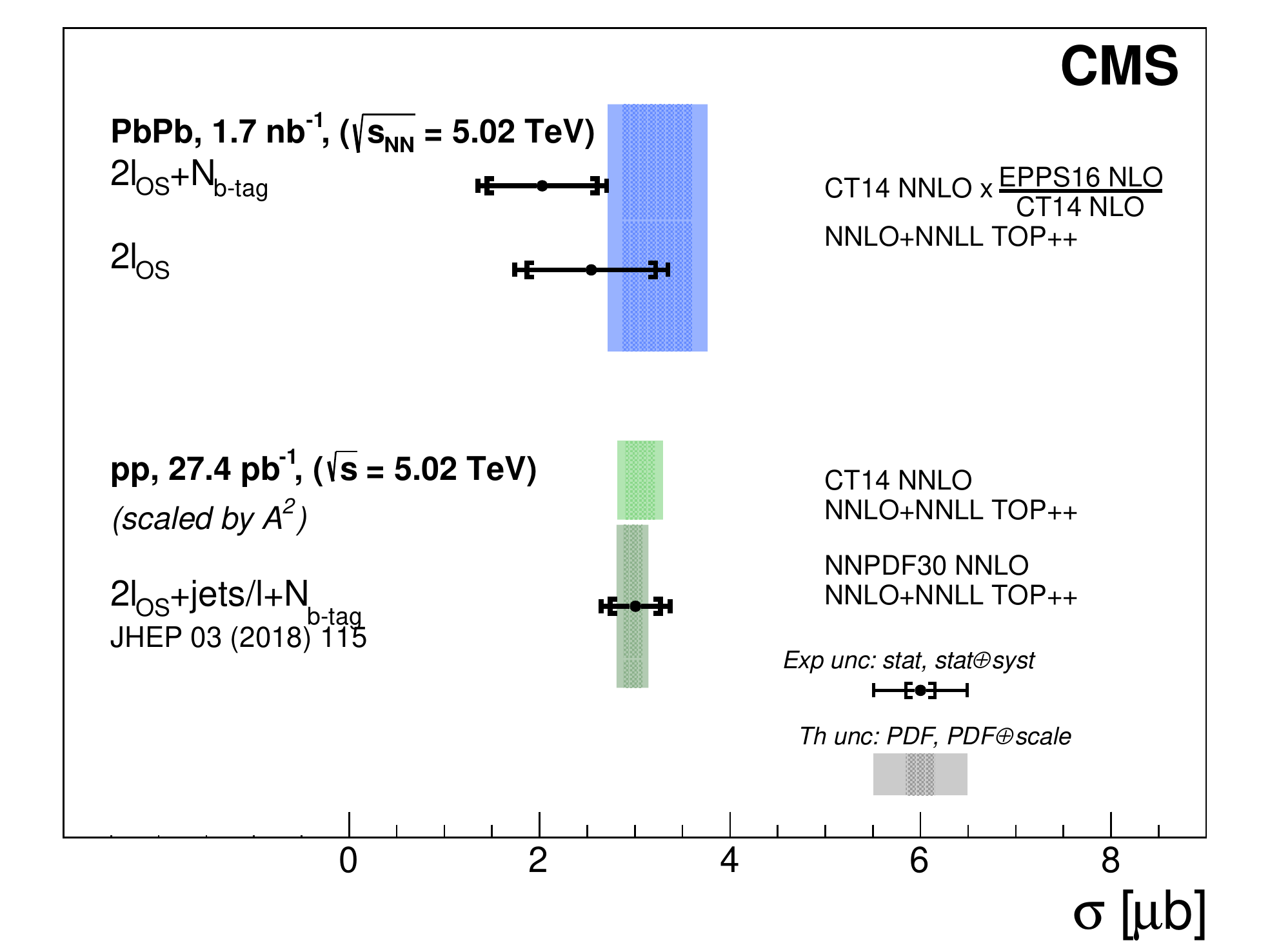}
\caption{Inclusive \ttbar\ cross sections measured with two methods in the combined \empe, \mmpm, and \empm\ final states in \PbPb\ collisions at $\sqrtsNN=5.02\TeV$, and \pp results at $\roots=5.02\TeV$ (scaled by $A^2$)
  from Ref.~\cite{Sirunyan:2017ule}. The measurements are compared with theoretical predictions at NNLO+NNLL accuracy in QCD~\cite{Czakon:2011xx,Czakon:2013goa}.
  The inner (outer) experimental uncertainty bars include statistical (statistical and systematic, added in quadrature) uncertainties.
  The inner (outer) theoretical uncertainty bands correspond to nuclear~\cite{Eskola:2016oht,Kusina:2016fxy} or free-nucleon~\cite{Dulat:2015mca,Ball:2014} PDF (PDF and scale, added in quadrature) uncertainties.
\label{fig:sqrt}}
\end{figure}

The inclusive \ttbar production cross sections (for the \methone\ and \methtwo\ methods) are finally obtained in the combined  \empe, \mmpm, and \empm\ final states multiplying
the best fit $\mu$ values of $\muobstwol$ and $\muobstwoljets$ by the theoretical expectation.
We measure
$\stt= \stttwol$ and $\stttwoljets$\mub for the two methods, \ie, smaller than, but still consistent with, the theoretical predictions at NNLO+NNLL accuracy in QCD.
Despite the expected antishadowing effect, the data appear below the theoretical expectations with or without nPDF effects.
Figure~\ref{fig:sqrt} presents a summary of the extracted cross sections, including the measurement in \pp collisions at
$\roots=5.02\TeV$~\cite{Sirunyan:2017ule} scaled  by $A^2$, compared with the corresponding theoretical predictions.

In summary, evidence for top quark pair (\ttbar) production is presented for the first time in nucleus-nucleus collisions,
irrespective of any possible final-state interactions of the studied top quark decay products (charged leptons and bottom quarks) with the quark-gluon plasma (QGP).
Using lead-lead collisions with a total integrated luminosity of \RecoLumi\ at a nucleon-nucleon center-of-mass energy of 5.02\TeV,
we measure the inclusive \ttbar\ cross section ($\stt$) utilizing the leptons only, and in a second method, in addition, the bottom quarks.
The extracted $\stt= \stttwol$ and $\stttwoljets$\mub in the two methods, respectively, are compatible with, though somewhat lower than, the expectations from scaled proton-proton data
and perturbative quantum chromodynamics calculations. This measurement is just the first step in using the top quark as a novel and powerful probe of the QGP.

\begin{acknowledgments}
  We congratulate our colleagues in the CERN accelerator departments for the excellent performance of the LHC and thank the technical and administrative staffs at CERN and at other CMS institutes for their contributions to the success of the CMS effort. In addition, we gratefully acknowledge the computing centers and personnel of the Worldwide LHC Computing Grid for delivering so effectively the computing infrastructure essential to our analyses. Finally, we acknowledge the enduring support for the construction and operation of the LHC and the CMS detector provided by the following funding agencies: BMBWF and FWF (Austria); FNRS and FWO (Belgium); CNPq, CAPES, FAPERJ, FAPERGS, and FAPESP (Brazil); MES (Bulgaria); CERN; CAS, MoST, and NSFC (China); COLCIENCIAS (Colombia); MSES and CSF (Croatia); RIF (Cyprus); SENESCYT (Ecuador); MoER, ERC IUT, PUT and ERDF (Estonia); Academy of Finland, MEC, and HIP (Finland); CEA and CNRS/IN2P3 (France); BMBF, DFG, and HGF (Germany); GSRT (Greece); NKFIA (Hungary); DAE and DST (India); IPM (Iran); SFI (Ireland); INFN (Italy); MSIP and NRF (Republic of Korea); MES (Latvia); LAS (Lithuania); MOE and UM (Malaysia); BUAP, CINVESTAV, CONACYT, LNS, SEP, and UASLP-FAI (Mexico); MOS (Montenegro); MBIE (New Zealand); PAEC (Pakistan); MSHE and NSC (Poland); FCT (Portugal); JINR (Dubna); MON, RosAtom, RAS, RFBR, and NRC KI (Russia); MESTD (Serbia); SEIDI, CPAN, PCTI, and FEDER (Spain); MOSTR (Sri Lanka); Swiss Funding Agencies (Switzerland); MST (Taipei); ThEPCenter, IPST, STAR, and NSTDA (Thailand); TUBITAK and TAEK (Turkey); NASU (Ukraine); STFC (United Kingdom); DOE and NSF (USA). \end{acknowledgments}

\bibliography{auto_generated}

\providecommand{\href}[2]{#2}\begingroup\raggedright\begin{thebibliography}{10}%
\makeatletter
\providecommand{\hrefCMSnoop }[0]{\@secondoftwo}%
\makeatother
\providecommand{\doi}{\texttt{doi:}\begingroup \urlstyle{tt}\Url}

\bibitem{Abe:1995hr}
\hrefCMSnoop {}{{CDF} Collaboration, ``Observation of top quark production in
  {\Pp{}\Pap} collisions'',} \textit{ Phys. Rev. Lett.} \textbf{ 74} (1995)
  2626,
  \href{http://dx.doi.org/10.1103/PhysRevLett.74.2626}{\doi{10.1103/PhysRevLett.74.2626}},
\href{http://www.arXiv.org/abs/hep-ex/9503002}{\texttt{arXiv:hep-ex/9503002}}.

\bibitem{D0:1995jca}
\hrefCMSnoop {}{{D0} Collaboration, ``Observation of the top quark'',} \textit{
  Phys. Rev. Lett.} \textbf{ 74} (1995) 2632,
  \href{http://dx.doi.org/10.1103/PhysRevLett.74.2632}{\doi{10.1103/PhysRevLett.74.2632}},
\href{http://www.arXiv.org/abs/hep-ex/9503003}{\texttt{arXiv:hep-ex/9503003}}.

\bibitem{Aad:2019hzw}
\hrefCMSnoop {}{{ATLAS Collaboration}, ``Measurement of the $\ttbar$ production
  cross section and lepton differential distributions in {\Pe\Pgm} dilepton
  events from $\pp$ collisions at $\sqrt{s}=13$ {\TeV} with the {ATLAS}
  detector'',} \textit{ Eur. Phys. J. C} \textbf{ 80} (2020) 528,
  \href{http://dx.doi.org/10.1140/epjc/s10052-020-7907-9}{\doi{10.1140/epjc/s10052-020-7907-9}},
\href{http://www.arXiv.org/abs/1910.08819}{\texttt{arXiv:1910.08819}}.

\bibitem{Aaij:2018imy}
\hrefCMSnoop {}{{LHCb Collaboration}, ``Measurement of forward top pair
  production in the dilepton channel in pp collisions at $\sqrt{s}$ = 13
  {\TeV}'',} \textit{ JHEP} \textbf{ 08} (2018) 174,
  \href{http://dx.doi.org/10.1007/JHEP08(2018)174}{\doi{10.1007/JHEP08(2018)174}},
\href{http://www.arXiv.org/abs/1803.05188}{\texttt{arXiv:1803.05188}}.

\bibitem{Sirunyan:2018goh}
\hrefCMSnoop {}{{CMS Collaboration}, ``Measurement of the $\ttbar$ production
  cross section, the top quark mass, and the strong coupling constant using
  dilepton events in pp collisions at $\sqrt{s}=13$ {\TeV}'',} \textit{ Eur.
  Phys. J. C} \textbf{ 79} (2019) 368,
  \href{http://dx.doi.org/10.1140/epjc/s10052-019-6863-8}{\doi{10.1140/epjc/s10052-019-6863-8}},
\href{http://www.arXiv.org/abs/1812.10505}{\texttt{arXiv:1812.10505}}.

\bibitem{Sirunyan:2017ule}
\hrefCMSnoop {}{{CMS Collaboration}, ``Measurement of the inclusive {\ttbar}
  cross section in pp collisions at $\sqrt{s} =$ 5.02 {\TeV} using final states
  with at least one charged lepton'',} \textit{ JHEP} \textbf{ 03} (2018) 115,
  \href{http://dx.doi.org/10.1007/JHEP03(2018)115}{\doi{10.1007/JHEP03(2018)115}},
\href{http://www.arXiv.org/abs/1711.03143}{\texttt{arXiv:1711.03143}}.

\bibitem{Khachatryan:2015fwh}
\hrefCMSnoop {}{{CMS Collaboration}, ``{Measurement of the $\ttbar$ production
  cross section in the all-jets final state in pp collisions at $\sqrt{s}=8$
  $\,{\TeV}$}'',} \textit{ Eur. Phys. J. C} \textbf{ 76} (2016) 128,
  \href{http://dx.doi.org/10.1140/epjc/s10052-016-3956-5}{\doi{10.1140/epjc/s10052-016-3956-5}},
\href{http://www.arXiv.org/abs/1509.06076}{\texttt{arXiv:1509.06076}}.

\bibitem{Sirunyan:2017xku}
\hrefCMSnoop {}{{CMS Collaboration}, ``Observation of top quark production in
  proton-nucleus collisions'',} \textit{ Phys. Rev. Lett.} \textbf{ 119} (2017)
  242001,
  \href{http://dx.doi.org/10.1103/PhysRevLett.119.242001}{\doi{10.1103/PhysRevLett.119.242001}},
\href{http://www.arXiv.org/abs/1709.07411}{\texttt{arXiv:1709.07411}}.

\bibitem{dEnterria:2015mgr}
\hrefCMSnoop {}{D.~d'Enterria, K.~Krajcz\'{a}r, and H.~Paukkunen, ``Top quark
  production in proton-nucleus and nucleus-nucleus collisions at {LHC} energies
  and beyond'',} \textit{ Phys. Lett. B} \textbf{ 746} (2015) 64,
  \href{http://dx.doi.org/10.1016/j.physletb.2015.04.044}{\doi{10.1016/j.physletb.2015.04.044}},
\href{http://www.arXiv.org/abs/1501.05879}{\texttt{arXiv:1501.05879}}.

\bibitem{Apolinario:2017sob}
\hrefCMSnoop {}{L.~Apolin\'{a}rio, J.~G. Milhano, G.~P. Salam, and C.~A.
  Salgado, ``Probing the time structure of the quark-gluon plasma with top
  quarks'',} \textit{ Phys. Rev. Lett.} \textbf{ 120} (2018) 232301,
  \href{http://dx.doi.org/10.1103/PhysRevLett.120.232301}{\doi{10.1103/PhysRevLett.120.232301}},
\href{http://www.arXiv.org/abs/1711.03105}{\texttt{arXiv:1711.03105}}.

\bibitem{CMS-PAS-LUM-17-002}
\href {http://cds.cern.ch/record/2628652}{{CMS Collaboration}, ``{CMS}
  luminosity measurement using 2016 proton-nucleus collisions at
  nucleon-nucleon center-of-mass energy of 8.16 {TeV}'',} CMS Physics Analysis
  Summary CMS-PAS-LUM-17-002, 2018.

\bibitem{Chatrchyan:2008zzk}
\hrefCMSnoop {}{{CMS Collaboration}, ``The {CMS} experiment at the {CERN}
  {LHC}'',} \textit{ JINST} \textbf{ 3} (2008) S08004,
\href{http://dx.doi.org/10.1088/1748-0221/3/08/S08004}{\doi{10.1088/1748-0221/3/08/S08004}}.

\bibitem{Czakon:2013tha}
\hrefCMSnoop {}{M.~Czakon, M.~L. Mangano, A.~Mitov, and J.~Rojo, ``{Constraints
  on the gluon PDF from top quark pair production at hadron colliders}'',}
  \textit{ JHEP} \textbf{ 07} (2013) 167,
  \href{http://dx.doi.org/10.1007/JHEP07(2013)167}{\doi{10.1007/JHEP07(2013)167}},
\href{http://www.arXiv.org/abs/1303.7215}{\texttt{arXiv:1303.7215}}.

\bibitem{Tanabashi:2018oca}
\hrefCMSnoop {}{{Particle Data Group}, M.~Tanabashi {et~al.}, ``Review of
  particle physics'',} \textit{ Phys. Rev. D} \textbf{ 98} (2018) 030001,
\href{http://dx.doi.org/10.1103/PhysRevD.98.030001}{\doi{10.1103/PhysRevD.98.030001}}.

\bibitem{Chatrchyan:2013exa}
\hrefCMSnoop {}{{CMS Collaboration}, ``Evidence of {\cPqb} jet quenching in
  {\PbPb} collisions at {$\sqrt{\smash[b]{s_{_{\mathrm{NN}}}}}=2.76$}
  {\TeV}'',} \textit{ Phys. Rev. Lett.} \textbf{ 113} (2014) 132301,
  \href{http://dx.doi.org/10.1103/PhysRevLett.113.132301}{\doi{10.1103/PhysRevLett.113.132301}},
  \href{http://www.arXiv.org/abs/1312.4198}{\texttt{arXiv:1312.4198}}.
[Erratum: \DOI{10.1103/PhysRevLett.115.029903}].

\bibitem{Khachatryan:2016bia}
\hrefCMSnoop {}{{CMS Collaboration}, ``The {CMS} trigger system'',} \textit{
  JINST} \textbf{ 12} (2017) P01020,
  \href{http://dx.doi.org/10.1088/1748-0221/12/01/P01020}{\doi{10.1088/1748-0221/12/01/P01020}},
\href{http://www.arXiv.org/abs/1609.02366}{\texttt{arXiv:1609.02366}}.

\bibitem{Sirunyan:2017ulk}
\hrefCMSnoop {}{{CMS Collaboration}, ``Particle-flow reconstruction and
  global-event description with the {CMS} detector'',} \textit{ JINST} \textbf{
  12} (2017) P10003,
  \href{http://dx.doi.org/10.1088/1748-0221/12/10/P10003}{\doi{10.1088/1748-0221/12/10/P10003}},
\href{http://www.arXiv.org/abs/1706.04965}{\texttt{arXiv:1706.04965}}.

\bibitem{Khachatryan:2015hwa}
\hrefCMSnoop {}{{CMS Collaboration}, ``Performance of electron reconstruction
  and selection with the {CMS} detector in proton-proton collisions at
  $\sqrt{s} = 8$ {\TeV}'',} \textit{ JINST} \textbf{ 10} (2015) P06005,
  \href{http://dx.doi.org/10.1088/1748-0221/10/06/P06005}{\doi{10.1088/1748-0221/10/06/P06005}},
\href{http://www.arXiv.org/abs/1502.02701}{\texttt{arXiv:1502.02701}}.

\bibitem{Sirunyan:2018fpa}
\hrefCMSnoop {}{{CMS Collaboration}, ``Performance of the {CMS} muon detector
  and muon reconstruction with proton-proton collisions at $\sqrt{s}=$ 13
  {\TeV}'',} \textit{ JINST} \textbf{ 13} (2018) P06015,
  \href{http://dx.doi.org/10.1088/1748-0221/13/06/P06015}{\doi{10.1088/1748-0221/13/06/P06015}},
\href{http://www.arXiv.org/abs/1804.04528}{\texttt{arXiv:1804.04528}}.

\bibitem{Cacciari:2011ma}
\hrefCMSnoop {}{M.~Cacciari, G.~P. Salam, and G.~Soyez, ``{FastJet} user
  manual'',} \textit{ Eur. Phys. J. C} \textbf{ 72} (2012) 1896,
  \href{http://dx.doi.org/10.1140/epjc/s10052-012-1896-2}{\doi{10.1140/epjc/s10052-012-1896-2}},
\href{http://www.arXiv.org/abs/1111.6097}{\texttt{arXiv:1111.6097}}.

\bibitem{Cacciari:2008gp}
\hrefCMSnoop {}{M.~Cacciari, G.~P. Salam, and G.~Soyez, ``The anti-$\kt$ jet
  clustering algorithm'',} \textit{ JHEP} \textbf{ 04} (2008) 063,
  \href{http://dx.doi.org/10.1088/1126-6708/2008/04/063}{\doi{10.1088/1126-6708/2008/04/063}},
\href{http://www.arXiv.org/abs/0802.1189}{\texttt{arXiv:0802.1189}}.

\bibitem{Cacciari:2007fd}
\hrefCMSnoop {}{M.~Cacciari and G.~P. Salam, ``Pileup subtraction using jet
  areas'',} \textit{ Phys. Lett. B} \textbf{ 659} (2008) 119,
  \href{http://dx.doi.org/10.1016/j.physletb.2007.09.077}{\doi{10.1016/j.physletb.2007.09.077}},
\href{http://www.arXiv.org/abs/0707.1378}{\texttt{arXiv:0707.1378}}.

\bibitem{Berta:2014eza}
\hrefCMSnoop {}{P.~Berta, M.~Spousta, D.~W. Miller, and R.~Leitner,
  ``Particle-level pileup subtraction for jets and jet shapes'',} \textit{
  JHEP} \textbf{ 06} (2014) 092,
  \href{http://dx.doi.org/10.1007/JHEP06(2014)092}{\doi{10.1007/JHEP06(2014)092}},
\href{http://www.arXiv.org/abs/1403.3108}{\texttt{arXiv:1403.3108}}.

\bibitem{Sirunyan:2017bsd}
\hrefCMSnoop {}{{CMS Collaboration}, ``Measurement of the splitting function in
  {\pp} and {\PbPb} collisions at {$\sqrt{\smash[b]{s_{_{\mathrm{NN}}}}} =$}
  5.02 {\TeV}'',} \textit{ Phys. Rev. Lett.} \textbf{ 120} (2018) 142302,
  \href{http://dx.doi.org/10.1103/PhysRevLett.120.142302}{\doi{10.1103/PhysRevLett.120.142302}},
\href{http://www.arXiv.org/abs/1708.09429}{\texttt{arXiv:1708.09429}}.

\bibitem{Sirunyan:2017ezt}
\hrefCMSnoop {}{{CMS Collaboration}, ``Identification of heavy-flavor jets with
  the {CMS} detector in {\pp} collisions at 13 {\TeV}'',} \textit{ JINST}
  \textbf{ 13} (2018) P05011,
  \href{http://dx.doi.org/10.1088/1748-0221/13/05/P05011}{\doi{10.1088/1748-0221/13/05/P05011}},
\href{http://www.arXiv.org/abs/1712.07158}{\texttt{arXiv:1712.07158}}.

\bibitem{Chatrchyan:2012nia}
\hrefCMSnoop {}{{CMS Collaboration}, ``{Jet momentum dependence of jet
  quenching in \PbPb collisions at $\sqrt{\smash[b]{s_{_{\mathrm{NN}}}}} =$
  2.76 \TeV}'',} \textit{ Phys. Lett. B} \textbf{ 712} (2012) 176,
  \href{http://dx.doi.org/10.1016/j.physletb.2012.04.058}{\doi{10.1016/j.physletb.2012.04.058}},
  \href{http://www.arXiv.org/abs/1202.5022}{\texttt{arXiv:1202.5022}}.

\bibitem{Alwall:2014hca}
J.~Alwall\hrefCMSnoop {}{ {et~al.}, ``The automated computation of tree-level
  and next-to-leading order differential cross sections, and their matching to
  parton shower simulations'',} \textit{ JHEP} \textbf{ 07} (2014) 079,
  \href{http://dx.doi.org/10.1007/JHEP07(2014)079}{\doi{10.1007/JHEP07(2014)079}},
\href{http://www.arXiv.org/abs/1405.0301}{\texttt{arXiv:1405.0301}}.

\bibitem{Sirunyan:2019dfx}
\hrefCMSnoop {}{{CMS Collaboration}, ``Extraction and validation of a new set
  of {CMS \PYTHIA8} tunes from underlying-event measurements'',} \textit{ Eur.
  Phys. J. C} \textbf{ 80} (2020) 4,
  \href{http://dx.doi.org/10.1140/epjc/s10052-019-7499-4}{\doi{10.1140/epjc/s10052-019-7499-4}},
\href{http://www.arXiv.org/abs/1903.12179}{\texttt{arXiv:1903.12179}}.

\bibitem{Sjostrand:2014zea}
T.~Sj{\"o}strand\hrefCMSnoop {}{ {et~al.}, ``An introduction to {\PYTHIA
  8.2}'',} \textit{ Comput. Phys. Commun.} \textbf{ 191} (2015) 159,
  \href{http://dx.doi.org/10.1016/j.cpc.2015.01.024}{\doi{10.1016/j.cpc.2015.01.024}},
\href{http://www.arXiv.org/abs/1410.3012}{\texttt{arXiv:1410.3012}}.

\bibitem{Re:2010bp}
\hrefCMSnoop {}{E.~Re, ``Single-top {\PW}{\PQt}-channel production matched with
  parton showers using the {\POWHEG} method'',} \textit{ Eur. Phys. J. C}
  \textbf{ 71} (2011) 1547,
  \href{http://dx.doi.org/10.1140/epjc/s10052-011-1547-z}{\doi{10.1140/epjc/s10052-011-1547-z}},
\href{http://www.arXiv.org/abs/1009.2450}{\texttt{arXiv:1009.2450}}.

\bibitem{Nason:2013ydw}
\hrefCMSnoop {}{P.~Nason and G.~Zanderighi, ``{\PWp{}\PWm, \PW{}\cPZ, and
  \cPZ{}\cPZ\ production in the \POWHEG-BOX-v2}'',} \textit{ Eur. Phys. J. C}
  \textbf{ 74} (2014) 2702,
  \href{http://dx.doi.org/10.1140/epjc/s10052-013-2702-5}{\doi{10.1140/epjc/s10052-013-2702-5}},
\href{http://www.arXiv.org/abs/1311.1365}{\texttt{arXiv:1311.1365}}.

\bibitem{Eskola:2016oht}
\hrefCMSnoop {}{K.~J. Eskola, P.~Paakkinen, H.~Paukkunen, and C.~A. Salgado,
  ``{EPPS16}: Nuclear parton distributions with {LHC} data'',} \textit{ Eur.
  Phys. J. C} \textbf{ 77} (2017) 163,
  \href{http://dx.doi.org/10.1140/epjc/s10052-017-4725-9}{\doi{10.1140/epjc/s10052-017-4725-9}},
\href{http://www.arXiv.org/abs/1612.05741}{\texttt{arXiv:1612.05741}}.

\bibitem{Dulat:2015mca}
S.~Dulat\hrefCMSnoop {}{ {et~al.}, ``New parton distribution functions from a
  global analysis of quantum chromodynamics'',} \textit{ Phys. Rev. D} \textbf{
  93} (2016) 033006,
  \href{http://dx.doi.org/10.1103/PhysRevD.93.033006}{\doi{10.1103/PhysRevD.93.033006}},
\href{http://www.arXiv.org/abs/1506.07443}{\texttt{arXiv:1506.07443}}.

\bibitem{Sirunyan:2018mlv}
\hrefCMSnoop {}{{CMS Collaboration}, ``Measurement of the top quark mass in the
  all-jets final state at $\sqrt{s} =$ 13 {TeV} and combination with the
  lepton+jets channel'',} \textit{ Eur. Phys. J. C} \textbf{ 79} (2019) 313,
  \href{http://dx.doi.org/10.1140/epjc/s10052-019-6788-2}{\doi{10.1140/epjc/s10052-019-6788-2}},
\href{http://www.arXiv.org/abs/1812.10534}{\texttt{arXiv:1812.10534}}.

\bibitem{CMSCollaboration:2015zni}
\href {http://cds.cern.ch/record/2020886}{{{CMS}} Collaboration, ``Technical
  proposal for the {Phase-II} upgrade of the {Compact Muon Solenoid}'',} CMS
  Technical proposal CERN-LHCC-2015-010, CMS-TDR-15-02, 2015.

\bibitem{Lokhtin:2005px}
\hrefCMSnoop {}{I.~P. Lokhtin and A.~M. Snigirev, ``A model of jet quenching in
  ultrarelativistic heavy ion collisions and high-\pt hadron spectra at
  {RHIC}'',} \textit{ Eur. Phys. J. C} \textbf{ 45} (2006) 211,
  \href{http://dx.doi.org/10.1140/epjc/s2005-02426-3}{\doi{10.1140/epjc/s2005-02426-3}},
\href{http://www.arXiv.org/abs/hep-ph/0506189}{\texttt{arXiv:hep-ph/0506189}}.

\bibitem{Agostinelli:2002hh}
\hrefCMSnoop {}{{GEANT4} Collaboration, ``{\GEANTfour}---a simulation
  toolkit'',} \textit{ Nucl. Instrum. Meth. A} \textbf{ 506} (2003) 250,
\href{http://dx.doi.org/10.1016/S0168-9002(03)01368-8}{\doi{10.1016/S0168-9002(03)01368-8}}.

\bibitem{Pedregosa:2012toh}
F.~Pedregosa\hrefCMSnoop {}{ {et~al.}, ``Scikit-learn: machine learning in
  {P}ython'',} \textit{ J. Mach. Learn. Res} \textbf{ 12} (2011) 2825,
  \href{http://dx.doi.org/10.5555/1953048.2078195}{\doi{10.5555/1953048.2078195}},
\href{http://www.arXiv.org/abs/1201.0490}{\texttt{arXiv:1201.0490}}.

\bibitem{Hocker:2007ht}
\hrefCMSnoop {}{H.~Voss, A.~H{\"o}cker, J.~Stelzer, and F.~Tegenfeldt,
  ``{TMVA}, the toolkit for multivariate data analysis with {ROOT}'',} in
  \textit{ XIth International Workshop on Advanced Computing and Analysis
  Techniques in Physics Research (ACAT)}, p.~40.
\newblock 2007.
\newblock
  \href{http://www.arXiv.org/abs/physics/0703039}{\texttt{arXiv:physics/0703039}}.
\newblock {[PoS(ACAT)040]}.
\href{http://dx.doi.org/10.22323/1.050.0040}{\doi{10.22323/1.050.0040}}.

\bibitem{Li:2012wna}
\hrefCMSnoop {}{Y.~Li and F.~Petriello, ``{Combining QCD and electroweak
  corrections to dilepton production in \FEWZ}'',} \textit{ Phys. Rev. D}
  \textbf{ 86} (2012) 094034,
  \href{http://dx.doi.org/10.1103/PhysRevD.86.094034}{\doi{10.1103/PhysRevD.86.094034}},
  \href{http://www.arXiv.org/abs/1208.5967}{\texttt{arXiv:1208.5967}}.

\bibitem{Campbell:2011bn}
\hrefCMSnoop {}{J.~M. Campbell, R.~K. Ellis, and C.~Williams, ``{Vector boson
  pair production at the LHC}'',} \textit{ JHEP} \textbf{ 07} (2011) 018,
  \href{http://dx.doi.org/10.1007/JHEP07(2011)018}{\doi{10.1007/JHEP07(2011)018}},
  \href{http://www.arXiv.org/abs/1105.0020}{\texttt{arXiv:1105.0020}}.

\bibitem{Kidonakis:2013zqa}
\hrefCMSnoop {}{N.~Kidonakis, ``Top quark production'',} in \textit{
  {Proceedings, Helmholtz International Summer School on Physics of Heavy
  Quarks and Hadrons (HQ 2013)}}, p.~139.
\newblock Verlag Deutsches Elektronen-Synchrotron, Hamburg, 2014.
\newblock \href{http://www.arXiv.org/abs/1311.0283}{\texttt{arXiv:1311.0283}}.
\newblock
\href{http://dx.doi.org/10.3204/DESY-PROC-2013-03/Kidonakis}{\doi{10.3204/DESY-PROC-2013-03/Kidonakis}}.

\bibitem{Verkerke:2005}
\hrefCMSnoop {}{W.~Verkerke and D.~P. Kirkby, ``{The RooFit toolkit for data
  modelling}'',} in \textit{ {Statistical Problems in Particle Physics,
  Astrophysics and Cosmology (PHYSTAT 05)}}, A.~H. Jaffe, ed.
\newblock World Scientific Publishing, Oxford University, UK, 2006.
\newblock
  \href{http://dx.doi.org/10.1142/9781860948985_0052}{\doi{10.1142/9781860948985_0052}}.

\bibitem{Moneta:2010pm}
L.~Moneta\hrefCMSnoop {}{ {et~al.}, ``The {RooStats} project'',} in \textit{
  {13th International Workshop on Advanced Computing and Analysis Techniques in
  Physics Research (ACAT2010)}}, p.~057.
\newblock Jaipur, India, 2010.
\newblock \href{http://www.arXiv.org/abs/1009.1003}{\texttt{arXiv:1009.1003}}.
\newblock {[PoS(ACAT2010)057]}.
\href{http://dx.doi.org/10.22323/1.093.0057}{\doi{10.22323/1.093.0057}}.

\bibitem{Khachatryan:2014jba}
\hrefCMSnoop {}{{CMS Collaboration}, ``Precise determination of the mass of the
  {Higgs} boson and tests of compatibility of its couplings with the standard
  model predictions using proton collisions at 7 and 8 {\TeV}'',} \textit{ Eur.
  Phys. J. C} \textbf{ 75} (2015) 212,
  \href{http://dx.doi.org/10.1140/epjc/s10052-015-3351-7}{\doi{10.1140/epjc/s10052-015-3351-7}},
\href{http://www.arXiv.org/abs/1412.8662}{\texttt{arXiv:1412.8662}}.

\bibitem{Cowan:2010js}
\hrefCMSnoop {}{G.~Cowan, K.~Cranmer, E.~Gross, and O.~Vitells, ``Asymptotic
  formulae for likelihood-based tests of new physics'',} \textit{ Eur. Phys. J.
  C} \textbf{ 71} (2011) 1554,
  \href{http://dx.doi.org/10.1140/epjc/s10052-011-1554-0}{\doi{10.1140/epjc/s10052-011-1554-0}},
  \href{http://www.arXiv.org/abs/1007.1727}{\texttt{arXiv:1007.1727}}.
[Erratum: \DOI{10.1140/epjc/s10052-013-2501-z}].

\bibitem{Czakon:2011xx}
\hrefCMSnoop {}{M.~Czakon and A.~Mitov, ``{\textsc{top++}}: a program for the
  calculation of the top pair cross section at hadron colliders'',} \textit{
  Comput. Phys. Commun.} \textbf{ 185} (2014) 2930,
  \href{http://dx.doi.org/10.1016/j.cpc.2014.06.021}{\doi{10.1016/j.cpc.2014.06.021}},
\href{http://www.arXiv.org/abs/1112.5675}{\texttt{arXiv:1112.5675}}.

\bibitem{Czakon:2013goa}
\hrefCMSnoop {}{M.~Czakon, P.~Fiedler, and A.~Mitov, ``Total top quark pair
  production cross section at hadron colliders through
  {$\mathcal{O}(\alpS^4)$}'',} \textit{ Phys. Rev. Lett.} \textbf{ 110} (2013)
  252004,
  \href{http://dx.doi.org/10.1103/PhysRevLett.110.252004}{\doi{10.1103/PhysRevLett.110.252004}},
\href{http://www.arXiv.org/abs/1303.6254}{\texttt{arXiv:1303.6254}}.

\bibitem{Ball:2014}
\hrefCMSnoop {}{{NNPDF} Collaboration, ``Parton distributions for the {LHC Run
  II}'',} \textit{ JHEP} \textbf{ 04} (2015) 040,
  \href{http://dx.doi.org/10.1007/JHEP04(2015)040}{\doi{10.1007/JHEP04(2015)040}},
\href{http://www.arXiv.org/abs/1410.8849}{\texttt{arXiv:1410.8849}}.

\bibitem{Arleo:2002kh}
\hrefCMSnoop {}{F.~Arleo, ``Tomography of cold and hot {QCD} matter: tools and
  diagnosis'',} \textit{ JHEP} \textbf{ 11} (2002) 044,
  \href{http://dx.doi.org/10.1088/1126-6708/2002/11/044}{\doi{10.1088/1126-6708/2002/11/044}},
\href{http://www.arXiv.org/abs/hep-ph/0210104}{\texttt{arXiv:hep-ph/0210104}}.

\bibitem{Arleo:2017ntr}
\hrefCMSnoop {}{F.~Arleo, ``Quenching of hadron spectra in heavy ion collisions
  at the {LHC}'',} \textit{ Phys. Rev. Lett.} \textbf{ 119} (2017) 062302,
  \href{http://dx.doi.org/10.1103/PhysRevLett.119.062302}{\doi{10.1103/PhysRevLett.119.062302}},
\href{http://www.arXiv.org/abs/1703.10852}{\texttt{arXiv:1703.10852}}.

\bibitem{Chatrchyan:2013iaa}
\hrefCMSnoop {}{{CMS Collaboration}, ``Measurement of {Higgs} boson production
  and properties in the {WW} decay channel with leptonic final states'',}
  \textit{ JHEP} \textbf{ 01} (2014) 096,
  \href{http://dx.doi.org/10.1007/JHEP01(2014)096}{\doi{10.1007/JHEP01(2014)096}},
\href{http://www.arXiv.org/abs/1312.1129}{\texttt{arXiv:1312.1129}}.

\bibitem{Khachatryan:2016kdb}
\hrefCMSnoop {}{{CMS Collaboration}, ``Jet energy scale and resolution in the
  {CMS} experiment in pp collisions at 8 {TeV}'',} \textit{ JINST} \textbf{ 12}
  (2017) P02014,
  \href{http://dx.doi.org/10.1088/1748-0221/12/02/P02014}{\doi{10.1088/1748-0221/12/02/P02014}},
\href{http://www.arXiv.org/abs/1607.03663}{\texttt{arXiv:1607.03663}}.

\bibitem{Kusina:2016fxy}
A.~Kusina\hrefCMSnoop {}{ {et~al.}, ``Vector boson production in {\pPb} and
  {\PbPb} collisions at the {LHC} and its impact on {nCTEQ15} {PDFs}'',}
  \textit{ Eur. Phys. J. C} \textbf{ 77} (2017) 488,
  \href{http://dx.doi.org/10.1140/epjc/s10052-017-5036-x}{\doi{10.1140/epjc/s10052-017-5036-x}},
\href{http://www.arXiv.org/abs/1610.02925}{\texttt{arXiv:1610.02925}}.

\end{thebibliography}\endgroup

\numberwithin{figure}{section}
\ifthenelse{\boolean{cms@external}}{}{
\clearpage\appendix
\section{\texorpdfstring{\ttbar}{ttbar} event display and BDT distributions in the \texorpdfstring{\empe}{empe}  and \texorpdfstring{\mmpm}{mmpm}  final states\label{app:suppMat}}

Dedicated algorithms deployed in real time allow the CMS detector to collect events with high-\pt leptons, hence making
the measurement of \ttbar\ production in \PbPb collisions possible in the \empe, \mmpm, and \empm\ final states.
Figure~\ref{fig:display} displays a candidate \ttbar\ event in the \empm\ final state in the \PbPb data sample.

\begin{figure*}[!htp]
\centering
\includegraphics[width=0.65\textwidth]{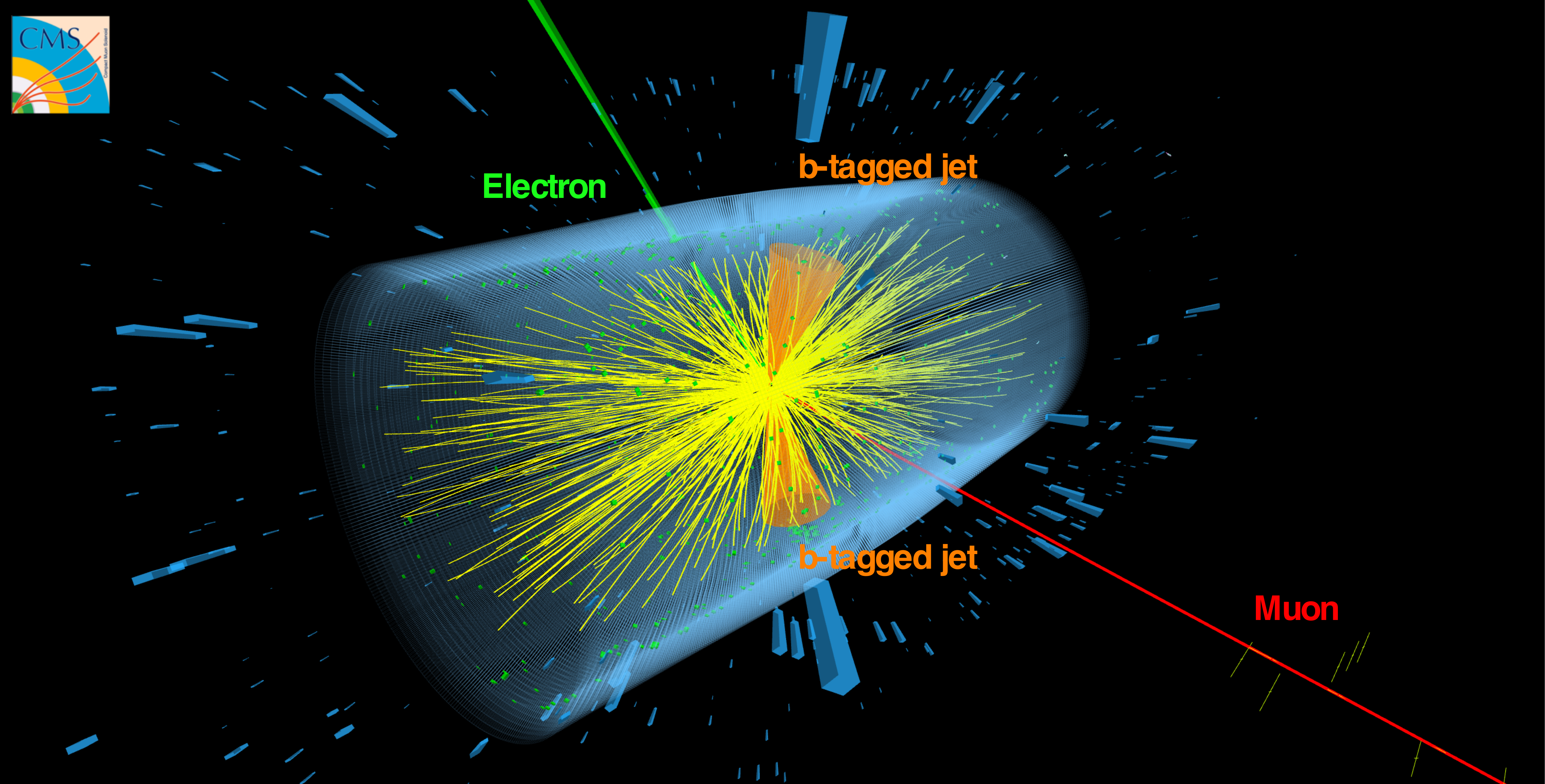}
\caption{Event display of a candidate \ttbar event measured in PbPb collisions where each top quark decays
 into a bottom quark and a \PW boson. The \PQb quarks and \PW bosons, in turn, produce jets and leptons, respectively.
The  event is interpreted as originating from the dilepton decay chain 
$\ttbar \to (\PQb \PWp) (\PAQb \PWm) \to (\PQb\, \Pep \PGne) (\PAQb\, \PGmm \PAGnGm)$.
\label{fig:display}}
\end{figure*}

The selected configuration for the multivariate analysis is a BDT with gradient boosting.
The classification probabilities for individual events are derived using a transformation of the background and signal distributions,
in which background events are uniformly distributed between 0 and 1, whereas signal  events  cluster  towards  1.
The expected BDT performance is evaluated by computing the area under the ``receiver operating characteristics'' curve,
yielding a value of 0.9 (an algorithm with ideal discrimination would yield 1.0, whereas with no discrimination would yield 0.5).
Cross validation with differently tuned parameters was performed, but no significant gain was observed.
Figures~\ref{fig:resultsBDT_sameflavor} and~\ref{fig:postfitdist_btag_sameflavor} show the observed BDT discriminator distributions for the \methone\
(as prefit expected) and \methtwo\ (as postfit predicted) methods, respectively, in the \empe (left) and \mmpm\ (right) final states.

\begin{figure*}[!htp]
    \centering
    \includegraphics[width=0.49\textwidth]{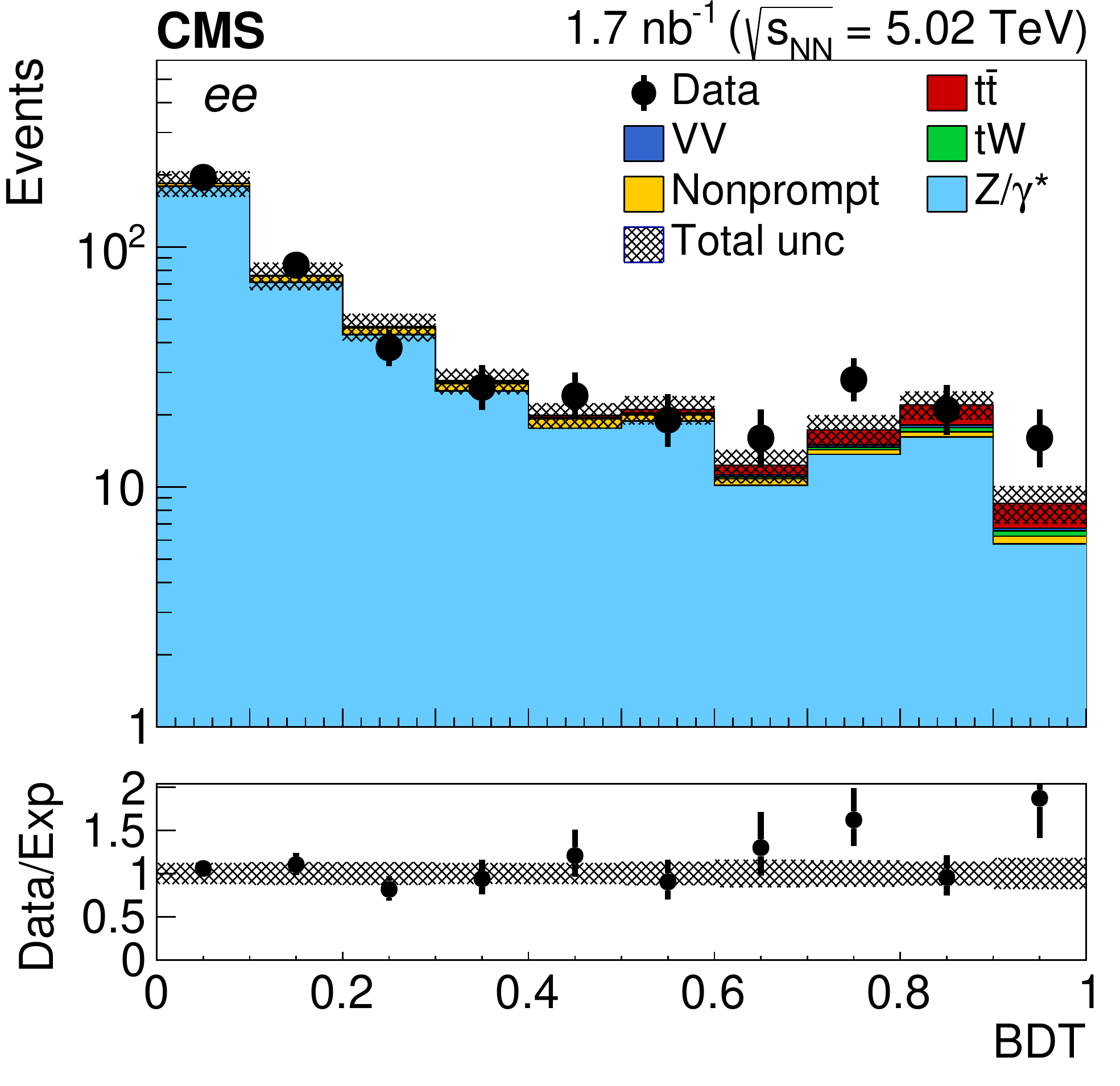}
    \includegraphics[width=0.49\textwidth]{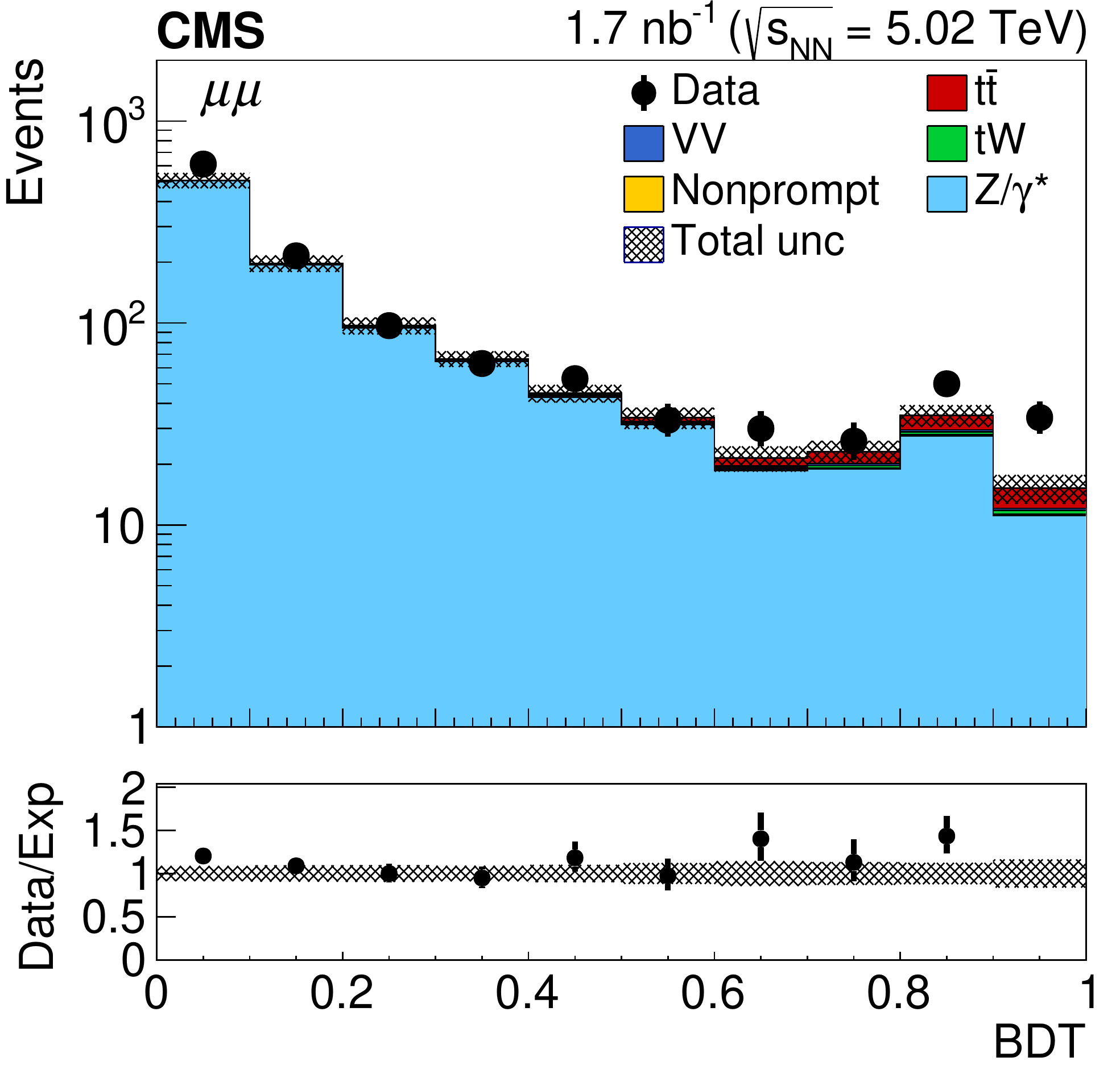}
    \caption{Observed (markers) and prefit expected (filled histograms) BDT discriminator distributions in the \empe\ (left) and \mmpm\ (right) final states.
    The data are shown with markers, and the signal and background processes with filled histograms.
    The vertical bars on the markers represent the statistical uncertainties in data. The hatched regions show the prefit uncertainties in the sum of \ttbar\ signal and backgrounds.
    The lower panels display the ratio of the data to expectations, including the \ttbar\ signal,  with bands representing the prefit uncertainties in the expectations.
}
    \label{fig:resultsBDT_sameflavor}
\end{figure*}

\begin{figure*}[!htbp]
    \centering
    \includegraphics[width=0.49\textwidth]{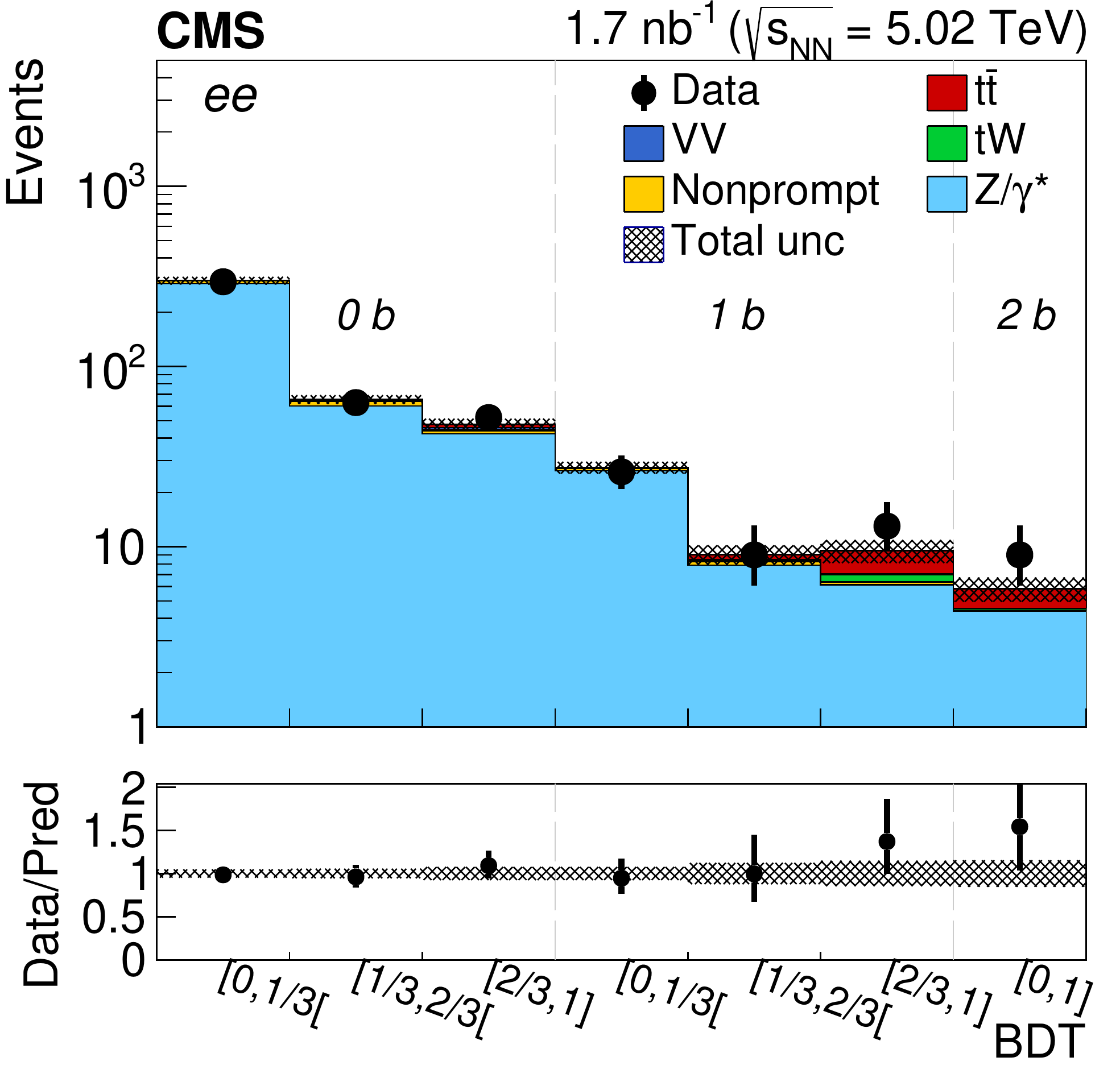}
    \includegraphics[width=0.49\textwidth]{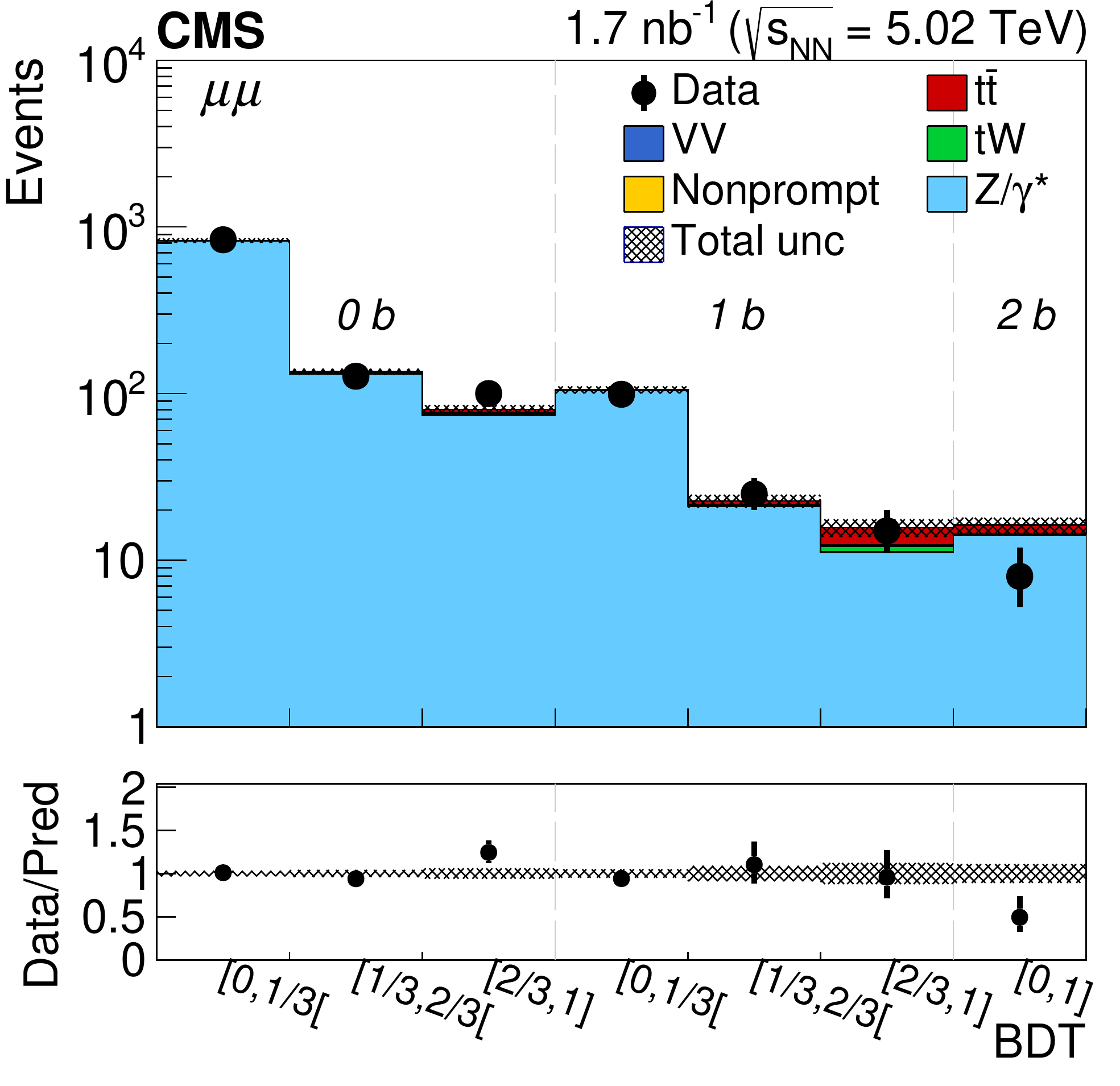}
    \caption{
    Observed (markers) and postfit predicted (filled histograms) BDT discriminator distributions in the \empe\ (left) and \mmpm\ (right) final states
    separately for the 0\PQb-, 1\PQb-, and 2\PQb-tagged jet  multiplicity categories.
    The data are shown with markers, and the signal and background processes with filled histograms.
    The vertical bars on the markers represent the statistical uncertainties in data. The hatched regions show the postfit uncertainties in the sum of \ttbar\ signal and backgrounds.
    The lower panels display the ratio of the data to predictions, including the \ttbar\ signal,  with bands representing the postfit uncertainties in the predictions.
      }
    \label{fig:postfitdist_btag_sameflavor}
\end{figure*}
}

\cleardoublepage \section{The CMS Collaboration \label{app:collab}}\begin{sloppypar}\hyphenpenalty=5000\widowpenalty=500\clubpenalty=5000\vskip\cmsinstskip
\textbf{Yerevan Physics Institute, Yerevan, Armenia}\\*[0pt]
A.M.~Sirunyan$^{\textrm{\dag}}$, A.~Tumasyan
\vskip\cmsinstskip
\textbf{Institut f\"{u}r Hochenergiephysik, Wien, Austria}\\*[0pt]
W.~Adam, F.~Ambrogi, T.~Bergauer, M.~Dragicevic, J.~Er\"{o}, A.~Escalante~Del~Valle, R.~Fr\"{u}hwirth\cmsAuthorMark{1}, M.~Jeitler\cmsAuthorMark{1}, N.~Krammer, L.~Lechner, D.~Liko, T.~Madlener, I.~Mikulec, N.~Rad, J.~Schieck\cmsAuthorMark{1}, R.~Sch\"{o}fbeck, M.~Spanring, S.~Templ, W.~Waltenberger, C.-E.~Wulz\cmsAuthorMark{1}, M.~Zarucki
% ...
\vskip\cmsinstskip
\dag: Deceased\\
1:  Also at Vienna University of Technology, Vienna, Austria\\
% ...
\end{sloppypar}
\end{document}